\documentclass[12pt,fleqn]{article}
\usepackage{graphicx,cite,color}

\def\theequation{\thesection.\arabic{equation}}
\newcommand{\newsection}[1]{\section{#1}\setcounter{equation}{0}}
\newcommand{\newappendix}[1]{\section*{#1}\setcounter{equation}{0}}

\def\be{\begin{equation}}
\def\ee{\end{equation}}
\def\bea{\begin{eqnarray}}
\def\eea{\end{eqnarray}}
\def\nnb{\nonumber}
\def\bbuildrel#1_#2^#3{\mathrel{\mathop{\kern 0pt#1}\limits_{#2}^{#3}}}
\def\slash#1{\setbox0=\hbox{$#1$}#1\hskip-\wd0\dimen0=5pt\advance
       \dimen0 by-\ht0\advance\dimen0 by\dp0\lower0.5\dimen0\hbox
         to\wd0{\hss\sl/\/\hss}}

\newcommand{\scs}{\scriptscriptstyle}
\newcommand{\f}{\frac}
\newcommand{\fm}[2]{{\textstyle \frac{#1}{#2}}}
\newcommand{\me}[1]{\langle#1\rangle}
\newcommand{\al}{\widetilde{\alpha}_{\mathrm s}}

\begin{document}

\begin{titlepage}

\begin{flushright}
TTP10-21\\
SFB/CPP-10-28\\
IFT-3/2010\\[3cm]
%arXiv:yymm.nnnn\\[2cm]
\end{flushright}

\begin{center}
\setlength {\baselineskip}{0.3in} 
{\bf\Large Large-{\boldmath $m_c$} Asymptotic Behaviour of\\[1mm]
  the {\boldmath ${\cal O}\left(\alpha_s^2\right)$}
  Corrections to {\boldmath $\bar B\to X_s\gamma$}}\\[3cm]
\setlength {\baselineskip}{0.2in}
{\large  Miko{\l}aj Misiak$^{1,2}$~ and~ Matthias Steinhauser$^1$}\\[5mm]
$^1$~{\it Institut f\"ur Theoretische Teilchenphysik, 
          Karlsruhe Institute of Technology (KIT),\\
          D-76128 Karlsruhe, Germany.}\\[5mm]
$^2$~{\it Institute of Theoretical Physics, University of Warsaw,\\
         Ho\.za 69, PL-00-681 Warsaw, Poland.}\\[3cm] 

{\bf Abstract}\\[5mm]
\end{center} 
\setlength{\baselineskip}{0.2in} 

We present details of our evaluation of the NNLO QCD corrections to~
${\cal B}(\bar B\to X_s\gamma)$~ in the heavy charm limit ($m_c \gg
\f{m_b}{2}$).  Results of this calculation have been essential for
estimating ${\cal O}\left(\alpha_s^2\right)$ effects in this branching
ratio via interpolation in $m_c$.

\end{titlepage}

\newsection{Introduction \label{sec:intro}}

The inclusive branching ratio ${\cal B}({\bar B}\to X_s\gamma)$ is well-known
to provide important constraints on extensions of the SM~\cite{Olive:2008vv}.
Its evaluation is based on an approximate equality of the hadronic and and
partonic decay widths\footnote{
  $\;$Corrections to Eq.~(\ref{main.approx}) of various origin have been widely
  discussed in the literature, most recently in Ref.~\cite{Benzke:2010js}. 
  A compact overview of previous results can be found in Ref.~\cite{Misiak:2009nr}.}
\be \label{main.approx}                                
\Gamma(\bar B \to X_s \gamma)_{{}_{E_\gamma > E_0}}
~\simeq~ \Gamma(b \to X_s^p \gamma)_{{}_{E_\gamma > E_0}},
\ee
where $X^p_s$ stands for $s$, $sg$, $sgg$, $sq\bar q$, etc. This
approximation works well only in a certain range of $E_0$, namely when
$E_0$ is large ($E_0 \sim \f{m_b}{2}$) but not too close to the endpoint
($m_b-2E_0\gg\Lambda_{\scs\rm QCD}$).  It has become customary to use
$E_0 = 1.6\;{\rm GeV} \simeq \f{m_b}{3}$ for comparing theory with
experiment.  Calculations including the ${\cal O}(\alpha_s^2)$ and
${\cal O}(\alpha_{\rm em})$ effects in the SM give~\cite{Misiak:2006zs,Misiak:2006ab}
\be \label{sm}
{\cal B}(\bar B \to X_s \gamma)_{{}_{E_\gamma > 1.6\,{\rm GeV}}}
= \left( 3.15 \pm 0.23 \right)\times 10^{-4},
\ee
where the uncertainty is dominated by ${\cal O}(\alpha_s \Lambda_{\scs\rm QCD}/m_b)$ 
non-perturbative effects~\cite{Benzke:2010js}.

The currently available experimental world averages read
\be \label{aver}
{\cal B}(\bar B \to X_s \gamma)_{{}_{E_\gamma > 1.6\,{\rm GeV}}} = \left\{ \begin{array}{l}
\left( 3.55 \pm 0.24_{\rm exp} \pm 0.09_{\rm model} \right)\times 10^{-4}~\mbox{\cite{Barberio:2008fa}},\\[1mm]
\left( 3.50 \pm 0.14_{\rm exp} \pm 0.10_{\rm model} \right)\times 10^{-4}~\mbox{\cite{Artuso:2009jw}}.
\end{array} \right.
\ee
They have been obtained from the measurements of CLEO~\cite{Chen:2001fj},
BABAR~\cite{Aubert:2005cu} and BELLE~\cite{Abe:2001hk,Limosani:2009qg} by
extrapolation in $E_0$ according to various photon energy spectrum models,
whose parameters have been fit to data.\footnote{
  $\;$Ref.~\cite{Barberio:2008fa} gives a larger error 
  than~\cite{Artuso:2009jw} because it includes results at $E_0 \geq
  1.8\;$GeV from the older measurements~\cite{Chen:2001fj,Aubert:2005cu}
  only, ignoring the more precise ones from
  Ref.~\cite{Limosani:2009qg}.}
The SM prediction (\ref{sm}) and the averages (\ref{aver}) are consistent
at the $1.2\sigma$ level.

The ${\cal O}(\alpha_s^2)$ contributions to the branching ratio amount to
around $10\%$, which exceeds the experimental errors and theoretical
non-perturbative uncertainties. However, these corrections have not been
included in a complete manner in Eq.~(\ref{sm}) because their charm-mass
dependence remains unknown beyond the
BLM-approximation~\cite{Brodsky:1982gc}. Instead, we have calculated all the
$m_c$-dependent non-BLM corrections in the~ $m_c \gg \f{m_b}{2}$~ limit, and
performed their interpolation in $m_c$ down to the measured value $m_c \simeq
\f{m_b}{4}$, assuming that they vanish at $m_c=0$. Our previous
paper~\cite{Misiak:2006ab} contains only the final analytic expressions for the
large-$m_c$ results together with a description of the interpolation.
Presenting details of the large-$m_c$ calculation is the purpose of the
present article.

The paper is organized as follows. Sec.~\ref{sec:notation} is devoted to
recalling the relevant definitions from
Ref~\cite{Misiak:2006ab}. Sec.~\ref{sec:exp} contains an explanation why
we did not use asymptotic expansions of three-loop on-shell Feynman
diagrams. Our actual method that involved charm decoupling at the
Lagrangian level is described in Secs.~\ref{sec:dec} and
\ref{sec:onshell}. We conclude in Sec.~\ref{sec:concl}.  Expressions for
the relevant functions $\phi^{(1)}_{ij}$ that originate from $b
\to s\gamma g$ are collected in the Appendix.

\newsection{Notation \label{sec:notation}}

We shall strictly follow the notation of Ref.~\cite{Misiak:2006ab}. The
present section collects the most important definitions only. The effective
Lagrangian that matters for evaluating QCD corrections to $b \to X_s^p \gamma$~ 
reads\footnote{
  $\;$Including the electroweak or CKM-suppressed corrections would require
  adding more terms to Eq.~(\ref{Leff5}).}
\be \label{Leff5}
{\cal L}_{\rm eff} ~=~ {\cal L}_{\scs {\rm QCD} \times {\rm QED}}(u,d,s,c,b) 
~+~ \f{4 G_F}{\sqrt{2}} V^*_{ts} V_{tb} \sum_{i=1}^{8} C_i Q_i,
\ee
where the local flavour-changing operators $Q_i$ arise from decoupling of the
$W$ boson and all the heavier particles. We shall need explicit expressions
for the following ones:
\bea
\begin{array}{rclcrcl}
Q_1 &=& (\bar{s}_L \gamma_{\mu} T^a c_L) (\bar{c}_L \gamma^{\mu} T^a b_L), &&
Q_2 &=& (\bar{s}_L \gamma_{\mu}     c_L) (\bar{c}_L \gamma^{\mu}     b_L),\\[2mm]
&&&&&& \hspace{-5cm} Q_4 ~=~ (\bar{s}_L \gamma_{\mu} T^a b_L) \sum_q (\bar{q}\gamma^{\mu} T^a q),\\[2mm]
Q_7 &=&  \f{e}{16\pi^2} m_b (\bar{s}_L \sigma^{\mu \nu}     b_R) F_{\mu \nu}, &&
Q_8 &=&  \f{g}{16\pi^2} m_b (\bar{s}_L \sigma^{\mu \nu} T^a b_R) G_{\mu \nu}^a.
\end{array}\nnb\\[-7mm] \label{ops}
\eea
The remaining three ($Q_3$, $Q_5$ and $Q_6$) are similar to $Q_4$ but
involve different Dirac and color structures. The sum over $q$ in $Q_4$
runs over all the active flavours ($u,d,s,c,b$). Masses of the light
quarks ($u,d,s$) are neglected throughout the paper, except for the 
collinear logarithm in Eq.~(\ref{phi88}) of the Appendix.

The Wilson Coefficients (WCs)~ $C_i$ are assumed to be $\overline{\rm
MS}$-renormalized at the scale $\mu_b \sim \f{m_b}{2}$. To avoid
scheme-dependence at the Leading Order (LO) in QCD, one usually works
with certain linear combinations called ``effective coefficients''
\be \label{ceffdef}
C_i^{\rm eff} = \left\{ \begin{array}{ll}
C_i, & \mbox{ for $i = 1, ..., 6$,} \\[1mm] 
C_7 + \sum_{j=1}^6 y_j C_j, & \mbox{ for $i = 7$,} \\[1mm]
C_8 + \sum_{j=1}^6 z_j C_j, & \mbox{ for $i = 8$,}
\end{array} \right.
\ee
where~ 
$y_j = (0, 0,-\f{1}{3}, -\f{4}{9}, -\f{20}{3}, -\f{80}{9})_j$~
and~ 
$z_j = (0, 0, 1, -\f{1}{6}, 20, -\f{10}{3})_j$~
in dimensional regularization with fully anticommuting $\gamma_5$.  In the SM,
all the $C_i^{\rm eff}(\mu_b)$ are known up to 
${\cal O}(\alpha_s^2)$~\cite{Misiak:2004ew,Gorbahn:2004my,Bobeth:1999mk}.

We are interested in evaluating the Next-to-Next-to-Leading Order (NNLO) QCD
corrections to the ratio of partonic radiative and charmless semileptonic decay rates
\be \label{pert.ratio}
\f{\Gamma[ b \to X_s \gamma]_{E_{\gamma} > E_0}}{
\Gamma[ b \to X_u e \bar{\nu}]} = 
\left| \f{ V^*_{ts} V_{tb}}{V_{ub}} \right|^2 
\f{6 \alpha_{\rm em}}{\pi} \sum_{i,j=1}^{8} C_i^{\rm eff}(\mu_b)\; C_j^{\rm eff}(\mu_b)\; K_{ij},
\ee
where the symmetric matrix $K_{ij}$ is perturbatively expanded as
\be
K_{ij} = \delta_{i7}\delta_{j7} + \al K_{ij}^{(1)} + \al^2 K_{ij}^{(2)} + {\cal O}\left(\al^3\right)
\hspace{2cm} \mbox{with} \hspace{1cm}
\al \equiv \f{\alpha_s^{(5)}(\mu_b)}{4 \pi}.
\ee  
Splitting $K_{ij}^{(2)}$ into the BLM ($K_{ij}^{(2)\beta_0}$) and non-BLM
($K_{ij}^{(2)\rm rem}$) pieces is done in a standard manner:
\be \label{ksplit}
K_{ij}^{(2)} ~=~ A_{ij}\, n_f + B_{ij} ~=~ K_{ij}^{(2)\beta_0} + K_{ij}^{(2)\rm rem}, 
\ee
where $n_f$ stands for the number of quark flavours in the effective
theory (\ref{Leff5}), and
\be
K_{ij}^{(2)\beta_0} \equiv -\f{3}{2}\beta_0 A_{ij} = -\f{3}{2} \left(11 - \f{2}{3} n_f\right) A_{ij},
\hspace{2cm}
K_{ij}^{(2)\rm rem} \equiv \f{33}{2} A_{ij} + B_{ij}.
\ee

In the ${\cal O}(\al^2)$ correction, one can safely ignore the small
$C^{(0)\rm eff}_{3,4,5,6}(\mu_b)$. Thus, it is sufficient to consider
$K_{ij}^{(2)}$ with $i,j\in\{1,2,7,8\}$ only. As far as
$K_{ij}^{(2)\beta_0}$ with such indices are concerned, all of them
except $K_{18}^{(2)\beta_0}$ and $K_{28}^{(2)\beta_0}$ are 
known~\cite{Bieri:2003ue,Boughezal:2007ny,Ligeti:1999ea,Ferroglia:2010xxx}
for the measured value of $m_c$. A calculation of $K_{18}^{(2)\beta_0}$ 
and $K_{28}^{(2)\beta_0}$ is underway~\cite{Misiak:2010xxx}.

Effects related to the absence of real $c\bar c$ production in $b \to
X_s^p \gamma$ and to non-zero masses of $b$ and $c$ quarks in
loops on gluon propagators belong to $K_{ij}^{(2)\rm rem}$.  They are
presently known for all the $i,j\in\{1,2,7,8\}$, either for arbitrary
$m_c$ or at least in the vicinity of its measured
value~\cite{Boughezal:2007ny,Asatrian:2006rq}. In fact, charm quark
loops on gluon propagators are the only source of $m_c$-dependent terms
in $K_{77}^{(2)}$, $K_{78}^{(2)}$ and $K_{88}^{(2)}$. Therefore, the
only quantities for which the $m_c$-interpolation still needs to be
performed are $K_{ij}^{(2)\rm rem}$ for $i\in\{1,2\}$ and
$j\in\{1,2,7,8\}$.  In the following, we shall restrict our
considerations to those cases only.
\begin{figure}[t]
  \begin{center}
    \includegraphics[width=15cm,angle=0]{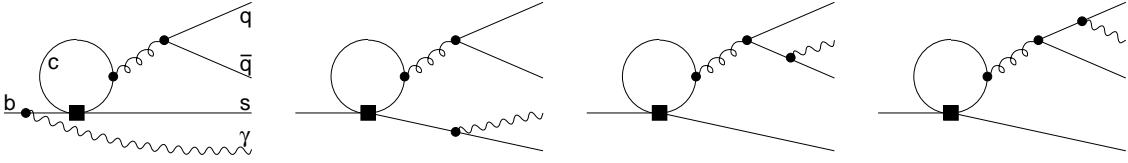}
    \caption{\sf $b \to s\gamma q\bar q$~ ($q=u,d,s$) diagrams 
                     with $Q_{1,2}$ vertices that survive in the large-$m_c$ limit.
      \label{fig:bsqqgam}}
  \end{center}
\end{figure}

Before closing this section, let us remark that our large-$m_c$
calculation is not 100\% complete.  There exist certain simple though
yet uncalculated contributions to $K_{ij}^{(2)\rm rem}$ with
$i\in\{1,2\}$ and $j\in\{1,2,7,8\}$ that survive in the
large-$m_c$ limit.  They originate from the four diagrams in
Fig.~\ref{fig:bsqqgam} that may interfere either with $b \to s\gamma
q\bar q$ contributions of $Q_{7,8}$ or just with themselves. Their
effect on the decay rate is of order $\al^2$, and it is expected to be
numerically very small due to limited four-body phase space left out by
the high photon energy cutoff $E_0 \sim \f{m_b}{3}$.  A convention
advocated in Ref.~\cite{Ligeti:1999ea}, which we follow here, is to
exclude those uncalculated terms from the BLM contribution, even though
some of them are proportional to the number of massless flavours.
We shall comment on this issue again in Sec.~\ref{sec:onshell}.

\newsection{Choice of the method\label{sec:exp}}

\begin{figure}[t]
  \begin{center}
    \includegraphics[width=8cm,angle=0]{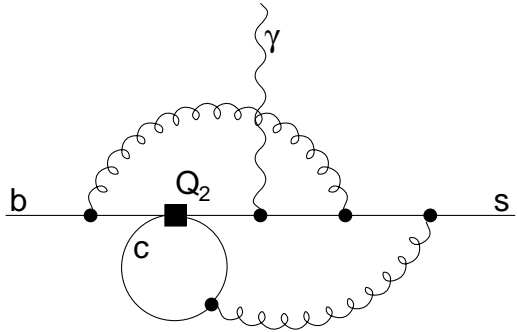}
    \caption{\sf One of the on-shell non-planar diagrams contributing to $K_{27}^{(2)\rm rem}$.
      \label{fig:nonplan}}
  \end{center}
\end{figure}

Our goal amounts to evaluating $K_{ij}^{(2)\rm rem}$ for $i\in\{1,2\}$
and $j\in\{1,2,7,8\}$ in the $m_c \gg \f{m_b}{2}$ limit, at the leading
order in $\f{m_b^2}{m_c^2}$.  On general grounds, one expects results of
the form
\be \label{asympk}
K_{ij}^{(2)\rm rem} = X^{(0)}_{ij} + X^{(1)}_{ij} \ln z + X^{(2)}_{ij} \ln^2 z +
{\cal O}\left(\f{1}{z}\right)
\hspace{1cm} \mbox{with} \hspace{1cm} z = \f{m_c^2}{m_b^2},
\ee 
where $X_{ij}^{(k)}$ are $m_c$-independent.  A straightforward method to
perform such a computation via asymptotic expansions~\cite{Smi02} would
involve three-loop on-shell vertex diagrams like the one shown in
Fig.~\ref{fig:nonplan} in the heavy-charm limit. 
Application of the so-called hard-mass procedure to such diagrams leads
to one-, two- and three-loop vacuum integrals with mass scale $m_c$, as
well as one- and two-loop on-shell vertex integrals with external
momenta $p_b^2=m_b^2$, $p_s^2=0$ and $p_\gamma^2=0$. Internal lines of
the vertex diagrams are either massless or carry mass $m_b$. At the
two-loop order many different cases occur since up to four bottom quark
lines can be present, and the photon can couple in all possible ways to
the charm or bottom quark.  In 2006, i.e. at the time when our actual
calculation~\cite{Misiak:2006ab} was performed, some of the relevant
two-loop on-shell massive vertex integrals remained unknown.
Furthermore, in addition to the virtual corrections, also contributions
from real gluon radiation had to be considered, which involved
phase-space integrals in parallel to the loop ones.

Although technical challenges related to the asymptotic expansion method
are certainly manageable, we decided to follow a field theory based
approach, as already mentioned in the Introduction.  This method takes
advantage of the fact that charm decoupling at the Lagrangian level can
be facilitated with the help of Equations of Motion (EOM). In effect,
all the necessary two-loop on-shell vertex integrals could be reduced to
the (planar) ones that are already known from the Next-to-Leading Order
(NLO) calculations of $b \to s \gamma$~\cite{Buras:2002tp}.

Before discussing in more detail the charm quark decoupling in the next
section, let us recall the large-$m_c$ results for $K_{ij}^{(2)\beta_0}$
with $i\in\{1,2\}$ and $j\in\{1,2,7,8\}$. Contributions to them from
the two-body channel $b \to s \gamma$ have been
evaluated with the help of asymptotic expansions using the program {\tt
exp}~\cite{Harlander:1997zb,Seidensticker:1999bb}.  Vacuum integrals
were treated with {\tt MATAD}\cite{Steinhauser:2000ry}, and the
reduction of two-loop vertex contributions to master integrals, which
can, e.g., be found in Ref.~\cite{Fleischer:1999tu}, was performed with
the help of {\tt AIR}~\cite{Anastasiou:2004vj}.  {\tt AIR} is written in
{\tt MAPLE} and is based on the Laporta
algorithm~\cite{Laporta:2001dd}. A more flexible and more powerful
alternative, which is available since 2008, would be the program {\tt
FIRE}~\cite{Smirnov:2008iw}.

Following Ref.~\cite{Misiak:2006ab}, we write
\mathindent0cm
\be \label{blm1}
  K_{27}^{(2)\beta_0} =   -6 K_{17}^{(2)\beta_0} = \beta_0\, 
  {\rm Re}\left\{ -\fm{3}{2} r_2^{(2)}(z) +2\left[a(z)+b(z)-\fm{290}{81}\right] L_b 
    - \fm{100}{81} L_b^2 \right\} + 2 \phi^{(2)\beta_0}_{27}(\delta),
\ee
\be \label{blm2}
  K_{ij}^{(2)\beta_0} = 2 \left(1+\delta_{ij}\right) \phi^{(2)\beta_0}_{ij}(\delta),
  \hspace{1cm} \mbox{for}~ i,j\neq 7,
\ee
\mathindent1cm
where~ $L_b = \ln \f{\mu_b^2}{m_b^2}$~ and~ $\delta = 1-\f{2E_0}{m_b}$.~
Our results for the first three terms of the large-$z$ expansion of
${\rm Re}\,r_2^{(2)}(z)$, ${\rm Re}\,a(z)$ and ${\rm Re}\,b(z)$ read
\bea
  {\rm Re}\,r_2^{(2)}(z) &=& \fm{8}{9}\ln^2 z + \fm{112}{243} \ln z + \fm{27650}{6561} 
  + \fm{1}{z} \left( \fm{38}{405} \ln^2 z - \fm{572}{18225} \ln z + \fm{10427}{30375}
    - \fm{8}{135} \pi^2 \right)\nnb\\[2mm] 
  &+& \fm{1}{z^2} \left( \fm{86}{2835} \ln^2 z - \fm{1628}{893025} \ln z +
    \fm{19899293}{125023500} - \fm{8}{405} \pi^2 \right) 
  + {\cal O}\left(\fm{1}{z^3}\right), \nnb\\[2mm]
{\rm Re}\,a(z) &=& \fm{4}{3} \ln z + \fm{34}{9}
+ \fm{1}{z} \left( \fm{5}{27} \ln z + \fm{101}{486} \right)
+ \fm{1}{z^2} \left( \fm{1}{15} \ln z + \fm{1393}{24300} \right)
+ {\cal O}\left(\fm{1}{z^3}\right), \nnb\\[2mm]
{\rm Re}\,b(z) &=& -\fm{4}{81} \ln z + \fm{8}{81}
- \fm{1}{z} \left( \fm{2}{45} \ln z + \fm{76}{2025} \right)
- \fm{1}{z^2} \left( \fm{4}{189} \ln z + \fm{487}{33075} \right)
+ {\cal O}\left(\fm{1}{z^3}\right),\mbox{~~} \label{asymp}
\eea
which has been confirmed in Ref.~\cite{Boughezal:2007ny} using a
    numerical evaluation of the coefficients at~ $z^{-k} \ln^n z$
    ($k,n=0,1,2$).~ The above functions are also known in the
    small-$z$ expansion~\cite{Bieri:2003ue,Greub:1996jd} -- see
    Eqs.~(3.9)--(3.10), (4.8) and Fig.~1 of Ref.~\cite{Misiak:2006ab}.

The functions $\phi^{(2)\beta_0}_{ij}(\delta)$ originate from the
bremsstrahlung~ $b \to s \gamma g$~ and~ $b \to s \gamma q\bar q$~
channels (but excluding the diagrams in Fig.~\ref{fig:bsqqgam}).  
Their numerical importance for~ $b \to X^p_s \gamma$~ is mild
due to the relatively high photon energy cutoff $E_0 \sim
\f{m_b}{3}$. At large $z$, all the $\phi^{(2)\beta_0}_{ij}(\delta)$ with
$i\in\{1,2\}$ and $j\in\{1,2,7,8\}$ behave as ${\cal
O}(\f{1}{z})$. Consequently, they can be ignored in the next section
where only the leading terms of the large-$z$ expansion of
$K^{(2)}_{ij}$ are considered.

\newsection{Charm decoupling: evaluation of the WCs\label{sec:dec}}

The matrices $X^{(n)}_{ij}$ in Eq.~(\ref{asympk}) can be obtained by
charm decoupling in the limit $m_c \gg \f{m_b}{2}$. In the first step,
we perform three-loop matching of the 5-flavour theory (\ref{Leff5}) onto
the 4-flavour one given by
\be \label{Leff4}
{\cal L}'_{\rm eff} ~=~ {\cal L}_{\scs {\rm QCD} \times {\rm QED}}(u,d,s,b) 
~+~ \fm{4 G_F}{\sqrt{2}} V^*_{ts} V_{tb} \sum_{i=3}^{8} C'_i Q'_i.
\ee
Here, $Q'_i$ differ from $Q_i$ in Eq.~(\ref{ops}) only by the absence of
charm-quark currents in $Q_{3,4,5,6}$. Additional terms containing
non-physical (evanescent and/or EOM-vanishing) operators on the
r.h.s. of Eqs.~(\ref{Leff5}) and (\ref{Leff4}) are implicitly assumed.
A complete list of such terms can be found in Sec.~3 of 
Ref.~\cite{Gorbahn:2004my}. Here, we just quote three examples of 
gauge-invariant EOM-vanishing operators
\mathindent0cm
\be \label{eomvan}
(\bar{s}_L \gamma^{\mu} T^a b_L) D^{\nu} G^a_{\mu \nu} +  g Q^{\scs (\;)}_4\!\!\!{}',
\hspace{6mm} \bar{s}_L \slash D \slash D \slash D b_L,\hspace{6mm} 
\f{ie}{16\pi^2} \left[ \bar{s}_L \stackrel{\leftarrow}{\slash D} \sigma^{\mu \nu} 
b_L F_{\mu \nu} - F_{\mu \nu} \bar{s}_L \sigma^{\mu \nu} \slash D b_L \right] + Q^{\scs (\;)}_7\!\!\!{}'\;.
\ee
\mathindent1cm

We proceed by analogy to our NNLO electroweak-scale matching for $b\to s
\gamma$ and $b\to s g$ \cite{Misiak:2004ew}. Requiring equality of
appropriately renormalized off-shell Green functions in both theories
leads to relations that allow to express $C'_i$ in terms of $C_i$.
Expansion in external momenta is performed prior to loop integration.
No IR regulators are introduced. All the particles except for the
decoupled charm are treated as massless, although linear terms in $m_b$
from bottom propagators and operator vertices are retained.  Diagrams like
the one in Fig.~\ref{fig:nonplan} still enter our calculation, but going off-shell
and expanding in external momenta makes their evaluation much easier.
The necessary three-loop integrals are found with the help of {\tt
MATAD}\cite{Steinhauser:2000ry}.

As well known, spurious IR divergences that appear in such a procedure
 (regulated dimensionally) cancel out in the final expressions for
 $C'_i$ in terms of $C_i$. All the loop diagrams that contain no charm
 quark are scaleless, so they vanish in dimensional
 regularization. Thus, $1/\epsilon^n$ poles on the 4-flavour theory side
 originate from the UV-renormalization constants only, while the
 5-flavour theory poles come from loop diagrams, too.

We identify the renormalization scale at which the matching is performed
with the previously introduced scale $\mu_b$. However, the charm mass
$m_c$ is assumed to be $\overline{\rm MS}$-renormalized at another scale
called $\mu_c$. The coefficients~ $C_i^{\rm eff}(\mu_b)$~ and~ $C'_i{}^{\rm
eff}(\mu_b)$~ are expanded in terms of~ $\al$~ and $\al'
=\fm{\alpha_s^{(4)}(\mu_b)}{4\pi}$,~ respectively, as follows:
\bea
C_i^{\rm eff}    &=& C^{(0)\rm eff}_i 
               + \al C^{(1)\rm eff}_i 
             + \al^2 C^{(2)\rm eff}_i + {\cal O}\left(\al^3\right), \nnb\\
C'_i{}^{\rm eff} &=& C'_i{}^{(0)\rm eff}
              + \al' C'_i{}^{(1)\rm eff} 
          + \al'{}^2 C'_i{}^{(2)\rm eff} + {\cal O}\left(\al'{}^3\right),
\eea
while $C'_i{}^{\rm eff}$ are related to $C'_i$ precisely as in
Eq.~(\ref{ceffdef}), with the same numbers $y_i$ and $z_i$.

Once the r.h.s. of Eq.~(\ref{pert.ratio}) is perturbatively expanded up
to ${\cal O}(\al^2)$, the sought $K^{(2)}_{ij}$ are multiplied by
$C_{1,2,7,8}^{(0)\rm eff}$ only. All the other unprimed WCs will be
set to zero everywhere in the following. After such a simplification,
our results for $C'_i{}^{\rm eff}$ take the form
\mathindent0cm
\be
C'_i{}^{(0)\rm eff} = C'_i{}^{(0)} = \left\{ \begin{array}{lcl}
0,                & \mbox{for} & i=3,4,5,6,\\[1mm]
C_i^{(0)\rm eff}, & \mbox{for} & i=7,8. \end{array}\right.\label{clo}
\ee
\be
C'_i{}^{(1)\rm eff} = \left\{ \begin{array}{lcl}
0,                & \mbox{for} & i=3,5,6,\\[1mm]
\fm{2}{3} \left(1-L_D\right)\left( C_2^{(0)} -\fm{1}{6} C_1^{(0)}\right) , & \mbox{for} & i=4,\\[1mm]
\left( \fm{218}{243} - \fm{208}{81} L_D \right) \left( C_2^{(0)} -\fm{1}{6} C_1^{(0)}\right), & \mbox{for} & i=7,\\[1mm]
\left( \fm{961}{3888} - \fm{173}{324} L_D \right) C_1^{(0)} 
+ \left(\fm{127}{324} - \fm{35}{27} L_D \right) C_2^{(0)}, & \mbox{for} & i=8. 
\end{array}\right.\label{cnlo}
\ee
\bea
C'_7{}^{(2)\rm eff} &=& 
\left[ \fm{2661293}{118098} - \fm{14293}{19683} n_\ell - \fm{3766}{729} \zeta_3
- \left( \fm{1877}{729} - \fm{220}{2187} n_\ell \right) L_D
+ \left( \fm{13384}{2187} - \fm{4}{27} n_\ell \right) L_D^2 
- \fm{832}{243} L_c \right] C_1^{(0)}\nnb\\[1mm] 
&+& \left[ - \fm{2861687}{19683} + \fm{28586}{6561} n_\ell + \fm{20060}{243} \zeta_3 
+ \left( \fm{2674}{243} -\fm{440}{729} n_\ell \right) L_D
- \left( \fm{15428}{729} - \fm{8}{9} n_\ell \right) L_D^2
+\fm{1664}{81} L_c \right] C_2^{(0)}\nnb\\[1mm]  
&+& \left(-\fm{364}{81} + \fm{112}{27} L_D - \fm{16}{9} L_D^2\right) C_7^{(0)\rm eff} + 
    \left( \fm{364}{243} - \fm{112}{81} L_D + \fm{16}{27} L_D^2 \right) C_8^{(0)\rm eff}.\label{cnnlo}
\eea
\mathindent1cm
where $L_D = \ln\f{\mu_b^2}{m_c^2}$, $L_c = \ln\f{\mu_c^2}{m_c^2}$, and
$n_\ell=3$ denotes the number of flavours that are kept massless
throughout the calculation. Retaining $n_\ell$ as a symbol is convenient for
cross-checking the BLM-part subtraction later on.

The above results have been obtained by matching~ $b \to s \gamma$,~ $b
\to s g$~ and~ $b \to sq\bar q$~ off-shell Green functions in both
theories. As a by-product, we have also obtained WCs of EOM-vanishing
operators like the ones in Eq.~(\ref{eomvan}). However, since on-shell
matrix elements of such operators vanish~\cite{Politzer:1980me}, there
is no need to consider them further.  This is precisely the point where
the Lagrangian-level decoupling is advantageous with respect to the
purely diagrammatic approach.  In the latter case, complicated on-shell
integrals may occur in contributions that are due to EOM-vanishing
operators alone, but this fact is not visible before reduction
to truly independent master integrals. An additional advantage in our
particular case is that we can use (in the next section) the known
two-loop on-shell~ $b\to s\gamma$~ matrix element of $Q_4$ that has been
evaluated without reduction to master integrals~\cite{Buras:2002tp}.

In the remainder of this section, let us recall several important points
concerning renormalization in off-shell matching calculations. First,
the external fields must be renormalized in an identical manner on both
sides of each matching equation.  One possibility is to renormalize all
the fields on shell. More conveniently, one can shift from the on-shell
to the $\overline{\rm MS}$ scheme on the 4-flavour theory side, and
perform an identical shift on the 5-flavour side. Second, one
adjusts the gauge coupling renormalization on the 5-flavour theory side
in such a way that the renormalized coupling equals to $\al'$ exactly
in~ $\epsilon = (4-D)/2$.~ At one loop, the necessary renormalization of
$\al'$ in the 5-flavour theory is given by~ $\al'{}^{\scs\rm BARE} =
Z_g^2 \al'$~ with
\be \label{zg}
Z_g = 1 ~+~ \al' \left[ -\f{25}{6} \left( \f{1}{\epsilon} -\gamma + \ln4\pi\right) ~+~ 
\f{\Gamma(1+\epsilon)}{3\epsilon} \left(\f{4\pi\mu_b}{m_c^2}\right)^\epsilon\; \right] 
~+~ {\cal O}(\al'{}^2),
\ee
in full analogy to Sec.~4 of Ref.~\cite{Misiak:2004ew} where more explanations 
can be found.\footnote{
$\;$Terms containing $(-\gamma+\ln 4\pi)$ are often skipped in Ref.~\cite{Misiak:2004ew}.
For more details about decoupling relations see Ref.~\cite{Chetyrkin:1997un}
and Eq.~(12) of Ref.~\cite{Grozin:2007fh}.}
Explicit expressions for shifts in the quark mass and wave function
renormalization can be found there, too.

As far as the WC renormalization is concerned 
 ($C_i^{\scs (\;)}\!\!\!{}' {}^{\;\scs\rm BARE} = \sum_j
   C_j^{\scs (\;)}\!\!\!{}'\,
Z_{ji}^{\scs (\;)}\!\!\!{}'\;$), 
we begin with the $\overline{\rm MS}$ scheme in both theories, and never
redefine the $Z_{ji}^{\scs (\;)}\!\!\!{}'\;$.  However, we re-express
them on the 5-flavour side in terms of $\al'$ that is
renormalized according to Eq.~(\ref{zg}).  This leads to appearance of
UV-finite terms in $Z_{ij}$ because the relation between $\al'$ and
$\al$ contains ${\cal O}(\epsilon)$ terms. Application of
D-dimensional rather than 4-dimensional relations between the gauge
couplings has been essential for successful tests of our expressions
(\ref{clo})--(\ref{cnnlo}) against results derived with the help of
asymptotic expansions in the off-shell case. These tests involved
direct three-loop $b\to s\gamma$ matching between the full SM and the
4-flavour effective theory~(\ref{Leff4}) for $m_c\ll M_W$,

\newsection{On--shell amplitudes \label{sec:onshell}}

We can now proceed to evaluating on-shell $b \to X_s \gamma$ amplitudes
in the 4-flavour theory using $C'_i{}^{(k)}$ as they stand in
Eqs.~(\ref{clo})--(\ref{cnnlo}). With all the gauge couplings factorized
out and the overall factor of~ $\f{4G_F}{\sqrt{2}} V_{ts}^* V_{tb}$~ omitted, 
the relevant expressions read
\mathindent0cm
\bea 
A(b \to s \gamma) &=& C'_7{}^{(0)} \me{Q'_7}^{(0)} ~+~
        \al' \left[ C'_7{}^{(1)\rm eff} \me{Q'_7}^{(0)} +
                    C'_7{}^{(0)} \me{Q'_7}^{(1)} +
                    C'_8{}^{(0)} \me{Q'_8}^{(1)} \right]\nnb\\[1mm] 
&+& \al'{}^2 \left[ C'_7{}^{(2)\rm eff} \me{Q'_7}^{(0)} +
                    C'_7{}^{(1)\rm eff} \me{Q'_7}^{(1)} +
                    C'_7{}^{(0)} \me{Q'_7}^{(2)} +
                    C'_8{}^{(1)\rm eff} \me{Q'_8}^{(1)}
\right. \nnb\\[1mm] &+& \left.
                    C'_8{}^{(0)} \me{Q'_8}^{(2)} +
                    C'_4{}^{(1)} \me{Q'_4}^{(2)}_{\rm eff} \right]
~+~ {\cal O}\left(\al'{}^3\right),\label{gam}\\[1mm]
A(b \to s \gamma g) &=& g'_s \left[ C'_7{}^{(0)} \me{Q'_7}^{(0)} + 
                                    C'_8{}^{(0)} \me{Q'_8}^{(0)} \right]
               ~+~ g'_s \al' \left[ C'_7{}^{(1)\rm eff} \me{Q'_7}^{(0)} + 
                                    C'_7{}^{(0)} \me{Q'_7}^{(1)}  
\right. \nnb\\[1mm] &+& \left.
                                    C'_8{}^{(1)\rm eff} \me{Q'_8}^{(0)} +
                                    C'_8{}^{(0)} \me{Q'_8}^{(1)} +
                                    C'_4{}^{(1)} \me{Q'_4}^{(1)}_{\rm eff} \right]
~+~ {\cal O}\left(g'_s \al'{}^2\right),\label{ggam}\\[1mm]
A(b \to s \gamma gg) &=& \al' \left[ C'_7{}^{(0)} \me{Q'_7}^{(0)} + 
                                     C'_8{}^{(0)} \me{Q'_8}^{(0)} \right]
~+~ {\cal O}\left(\al'{}^2\right),\label{gggam}\\[1mm]
A(b \to s \gamma q\bar q) &=& \al' \left[ C'_7{}^{(0)} \me{Q'_7}^{(0)} + 
                                          C'_8{}^{(0)} \me{Q'_8}^{(0)} +
                                          C'_4{}^{(1)} \me{Q'_4}^{(0)} \right]
~+~ {\cal O}\left(\al'{}^2\right),\label{qqgam}
\eea
\mathindent1cm
where $\me{Q'_j}^{(n)}$ denotes the $n$-loop renormalized matrix element of $Q'_j$
between the considered external states, and $C'_i{}^{(k)}$ are used only when
$C'_i{}^{(k)}=C'_i{}^{(k)\rm eff}$.

For the penguin operators (j=3,4,5,6) we have~ 
   $\me{Q_j^{\scs (\;)}\!\!\!{}'\,}^{(n)}_{\rm eff} \equiv 
    \me{Q_j^{\scs (\;)}\!\!\!{}'\,}^{(n)} - 
y_j \me{Q_7^{\scs (\;)}\!\!\!{}'\,}^{(n-1)} - 
z_j \me{Q_8^{\scs (\;)}\!\!\!{}'\,}^{(n-1)}$
with the same numbers $y_j$ and $z_j$ as in Eq.~(\ref{ceffdef}). In fact, those numbers
are determined by the requirement~
$\me{s\gamma|Q_j^{\scs (\;)}\!\!\!{}'\,|b}^{(1)}_{\rm eff} = 0 = 
 \me{sg     |Q_j^{\scs (\;)}\!\!\!{}'\,|b}^{(1)}_{\rm eff}$.~
Expressing amplitudes in terms of 
$\me{Q_j^{\scs (\;)}\!\!\!{}'\,}^{(n)}_{\rm eff}$
is convenient in other cases, too. For instance,
the bremsstrahlung matrix elements 
$\me{s\gamma g|Q_j^{\scs (\;)}\!\!\!{}'\,|b}^{(1)}_{\rm eff}$
are given by subsets of diagrams contributing to 
$\me{s\gamma g|Q_j^{\scs (\;)}\!\!\!{}'\,|b}^{(1)}$,
namely those where both the photon and the gluon are attached to the
quark loop~\cite{Pott:1995if}. At two loops, 
$\;\me{s\gamma|Q_j^{\scs (\;)}\!\!\!{}'\,|b}^{(2)}_{\rm eff}$
contain no IR divergences, contrary to
$\;\me{s\gamma|Q_j^{\scs (\;)}\!\!\!{}'\,|b}^{(2)}$~\cite{Buras:2002tp}.

In Eq.~(\ref{gam}), we need
\be 
\me{s\gamma|Q'_4\,|b}^{(2)}_{\rm eff} ~=~ \me{s\gamma|Q'_7\,|b}^{(0)} \left[ 
r'_4 + \left( -\f{20}{243} + \f{8}{81} n_\ell \right) L_b \right],\label{r4}
\ee
where
\be
{\rm Re}\,r'_4 ~=~ -\f{137}{729} - \f{52}{243} n_\ell - \f{4\pi}{9\sqrt{3}} - \f{16}{27} X_b 
+ \f{1}{6} {\rm Re}\,a(1) + \f{5}{3} {\rm Re}\,b(1), 
\ee
and
\mathindent0cm
\bea
X_b &=& \int_0^1 dx \int_0^1 dy \int_0^1 dv \; x y 
\ln\left[v + x (1-x) (1-v) (1-v + v y)\right] ~\simeq~ -0.1684, \nnb\\[3mm]
{\rm Re}\,a(1) &=& \fm{43}{9} + \fm{8}{9} \int_0^1 dx \int_0^1 dy \int_0^1 dv 
\left\{  [2 - v + xy (2v-3)] \ln [ v + x (1-x) (1-v) (1 - v + v y)]  
\right. \nonumber\\[2mm] &+& \left. 
[1-v+xy(2v-1)] \ln [ 1 - x(1-x)yv] \right\} ~\simeq~ 4.0859, \nnb\\[4mm]
{\rm Re}\,b(1) &=& \f{320}{81} - \f{4 \pi}{3 \sqrt{3}} 
+ \f{632}{1215} \pi^2 - \f{8}{45} \left[ \f{d^2 \ln \Gamma(x)}{dx^2} \right]_{x=\f{1}{6}} 
~\simeq~ 0.0316. \label{integs}
\eea
\mathindent1cm
The result in Eq.~(\ref{r4}) has been extracted from Eqs.~(3.1) and
(6.21) of Ref.~\cite{Buras:2002tp} after reintroducing explicit
$n_\ell$-dependence there. More precisely, setting $z=0$ in the quoted
equations of Ref.~\cite{Buras:2002tp} gives the same number for~ $r_4 +
\gamma_{47}^{(0)\rm eff} \ln(m_b/\mu_b)$~ there as setting $n_\ell =
4$ in the square bracket of Eq.~(\ref{r4}) here.  However, here we need
$n_\ell = 3$.\newpage

The next steps to perform are as follows:
\begin{itemize}
\item{} Calculate moduli squared of the amplitudes in Eqs.~(\ref{gam})--(\ref{qqgam}), 
        sum over polarizations and integrate over the phase space.
\item{} Re-expand everything in terms of~ $\al$~ using~ 
        $\al' = \al \left( 1 - \f{2}{3} \al L_D + {\cal O}(\al^2) \right)$,~ 
        and take into account normalization to the semileptonic rate in Eq.~(\ref{pert.ratio}).
\item{} Pick up only those ${\cal O}(\al^2)$ terms that contain at least a single $C_{1,2}^{(0)}$,
        and read out the corresponding $K_{ij}^{(2)}$.
\item{} Subtract the BLM contributions (leading terms in the large-$z$ expansion only) 
        using Eqs.~(\ref{blm1})--(\ref{blm2}) with $\beta_0 = 11 - \fm{2}{3}(n_\ell + 2)$, 
        and check that all the $n_\ell$-terms cancel out.
\end{itemize}

A brief look at Eqs.~(\ref{clo})--(\ref{cnlo}) and
(\ref{gam})--(\ref{qqgam}) ensures that the matrix element in
Eq.~(\ref{r4}) is actually the only two-loop on-shell one that we need.
Let us stress again that this is the case only thanks to identifying the 
EOM-vanishing operators at the Lagrangian level.

Another straightforward observation is that~ $K_{ij}^{(2)}$~ for~
$i,j\in\{1,2\}$~ receive contributions only from the~ $C'_7{}^{(1)\rm
eff} \me{Q'_7}^{(0)}$~ term in Eq.~(\ref{gam}) and the ~$C'_4{}^{(1)}
\me{Q'_4}^{(0)}$~ term in Eq.~(\ref{qqgam}). As the latter term is not
known, we shall neglect it in what follows.\footnote{ 
$\;$It originates precisely from the diagrams in Fig.~\ref{fig:bsqqgam} that
were discussed at the end of Sec.~\ref{sec:notation}. Now it is clear
that including their $n_\ell$-parts in the BLM approximation would not be
mandatory because the $q\bar q$ pair emitted from $Q'_4$ has no gluonic
counterpart in any other diagram. More precisely, this counterpart occurs
only in the first EOM-vanishing operator in Eq.~(\ref{eomvan}) that gives no
contribution on shell.}
With this approximation, everything we need is given by the quantity
that multiplies~ $\left( C_2^{(0)} -\fm{1}{6} C_1^{(0)}\right)$~ in~ $C'_7{}^{(1)\rm
eff}$~ (see Eq.~(\ref{cnlo})). It contains no $n_\ell$-piece, and
is equal to~ $K_{27}^{(1)} + {\cal O}(1/z)$. Consequently
\bea 
K_{22}^{(2)\rm rem} &=&
36\,K_{11}^{(2)\rm rem} + {\cal O}\left(\fm{1}{z}\right) ~=~
-6\,K_{12}^{(2)\rm rem} + {\cal O}\left(\fm{1}{z}\right) ~=~
\left(K_{27}^{(1)}\right)^2 + {\cal O}\left(\fm{1}{z}\right) \nnb\\[2mm]
&=& \left[ \fm{218}{243} - \fm{208}{81} L_D\right]^2 
+ {\cal O}\left(\fm{1}{z}\right), \label{k22rem}
\eea
which is identical to Eq.~(5.4) of Ref.~\cite{Misiak:2006ab}. 

It remains to determine $K_{(12)(78)}^{(2)\rm rem}$,~ i.e.,~
$K_{ij}^{(2)\rm rem}$ for $i\in\{1,2\}$ and $j\in\{7,8\}$. Once
$\me{s\gamma q\bar q|Q'_4|b}^{(0)}$ has been neglected, the only
relevant processes are~ $b \to s \gamma$~ and $b \to s \gamma g$,~ and
we need at least one beyond-LO coefficient $C'_i{}^{(k)\rm eff}$, which
makes the calculation very similar to the~ $b \to X^p_s \gamma$~ one at
the NLO.  In fact, all the necessary matrix elements and phase-space
integrals come in the same combinations as in the known results for
\bea
K'{}^{(1)}_{\!\!77} &=& K_{77}^{(1)} ~=~ 
-\fm{182}{9} + \fm{8}{9}\pi^2 - \fm{32}{3} L_b + 4\,\phi_{77}^{(1)}(\delta),\nnb\\[1mm]
K'{}^{(1)}_{\!\!78} &=& K_{78}^{(1)} ~=~ 
\fm{44}{9} - \fm{8}{27}\pi^2 +\fm{16}{9} L_b  + 2\,\phi_{78}^{(1)}(\delta),\nnb\\[1mm]
K'{}^{(1)}_{\!\!88} &=& K_{88}^{(1)} ~=~ 4\,\phi^{(1)}_{88}(\delta),\nnb\\[1mm]
K'{}^{(1)}_{\!\!47} &=& K_{47}^{(1)} - 2\, {\rm Re}\, b(z) + \fm{52}{243} - \fm{8}{81} L_b ~=~ 
{\rm Re}\, r'_4 + \left( -\fm{20}{243} + \fm{8}{81} n_\ell \right) L_b + 2\,\phi^{(1)}_{47}(\delta),\nnb\\[1mm] 
K'{}^{(1)}_{\!\!48} &=& K_{48}^{(1)} ~=~ 2\,\phi^{(1)}_{48}(\delta).\label{k1}
\eea
For completeness, all the relevant functions $\phi^{(1)}_{ij}(\delta)$
are collected in the Appendix.  Normalization to the semileptonic rate is
already taken into account in $K^{(1)}_{77}$, and there is no other point
where it could matter in the evaluation of $K_{(12)(78)}^{(2)\rm rem}$.

Once the quantities from Eq.~(\ref{k1}) are used, equations that
determine the sought $K^{(2)}_{ij}$ take a simple form
\mathindent7mm
\bea
C_1^{(0)} K_{17}^{(2)} + C_2^{(0)} K_{27}^{(2)} + {\cal O}\left(\fm{1}{z}\right)
&=& \widetilde{C}'_7{}^{(2)\rm eff} 
+ \left( K_{77}^{(1)} -\fm{2}{3} L_D \right) C'_7{}^{(1)\rm eff} 
+ K_{78}^{(1)} C'_8{}^{(1)\rm eff} + K'{}^{(1)}_{\!\!47} C'_4{}^{(1)}\nnb\\[1mm]
C_1^{(0)} K_{18}^{(2)} + C_2^{(0)} K_{28}^{(2)} + {\cal O}\left(\fm{1}{z}\right)
&=& K_{78}^{(1)} C'_7{}^{(1)\rm eff} + K_{88}^{(1)} C'_8{}^{(1)\rm eff} 
    + K_{48}^{(1)} C'_4{}^{(1)},\label{extract.k2}
\eea
where $\widetilde{C}'_7{}^{(2)\rm eff}$ stands for $C'_7{}^{(2)\rm
eff}$~(\ref{cnnlo}) with $C_{7,8}^{(0)\rm eff}$-terms set to zero.

In the last step, as already mentioned above, we need to subtract the
BLM parts (\ref{blm1})--(\ref{blm2}) from the calculated $K^{(2)}_{ij}$
to obtain $K^{(2)\rm rem}_{ij}$. At this level, it is convenient to
express $K'{}^{(1)}_{\!\!47}$ first in terms of $K_{47}^{(1)}$, and next
in terms of
\mathindent0cm
\bea
K_{47}^{(1)\rm rem} &=& K_{47}^{(1)} - \beta_0 \left( \fm{26}{81} - \fm{4}{27} L_b \right)\nnb\\[1mm]
&=& -\fm{2411}{729} - \fm{4\pi}{9\sqrt{3}} - \fm{16}{27} X_b 
+ \fm{1}{6} {\rm Re}\,a(1) + \fm{5}{3} {\rm Re}\,b(1) +
\fm{328}{243} L_b + \fm{8}{81} L_D + 2\,\phi^{(1)}_{47}(\delta)
+ {\cal O}\left(\fm{1}{z}\right).\nnb\\[1mm] \\[-8mm]\nnb
\eea
\mathindent1cm
Our final results for $K_{(12)(78)}^{(2)\rm rem}$ take the form\\[-2mm]
\bea 
K_{27}^{(2)\rm rem} &=& K_{27}^{(1)} K_{77}^{(1)} 
+ \left( \fm{127}{324} - \fm{35}{27} L_D \right) K_{78}^{(1)} 
+ \fm{2}{3} (1 - L_D ) K_{47}^{(1)\rm rem}
\nnb\\[2mm] &-&
\fm{4736}{729} L_D^2 + \fm{1150}{729} L_D - \fm{1617980}{19683} + \fm{20060}{243} \zeta_3
+ \fm{1664}{81} L_c + {\cal O}\left(\fm{1}{z}\right),\\[2mm]
K_{28}^{(2)\rm rem} &=& K_{27}^{(1)} K_{78}^{(1)} 
+ \left( \fm{127}{324} - \fm{35}{27} L_D \right) K_{88}^{(1)} 
+ \fm{2}{3} (1 - L_D ) K_{48}^{(1)}  + {\cal O}\left(\fm{1}{z}\right),\\[2mm]
K_{17}^{(2)\rm rem} &=&\!\! -\fm{1}{6} K_{27}^{(2)\rm rem} + \left(\fm{5}{16}-\fm{3}{4}L_D\right)K_{78}^{(1)}
-\fm{1237}{729} +\fm{232}{27}\zeta_3 +\fm{70}{27}L_D^2 -\fm{20}{27} L_D
+ {\cal O}\left(\fm{1}{z}\right)\!,\\[2mm]
K_{18}^{(2)\rm rem} &=&\!\! -\fm{1}{6} K_{28}^{(2)\rm rem} + \left(\fm{5}{16}-\fm{3}{4}L_D\right)K_{88}^{(1)}
+ {\cal O}\left(\fm{1}{z}\right),\\[-4mm]\nnb
\eea
\mathindent1cm
which is identical to Eqs.~(5.5)--(5.8) of Ref.~\cite{Misiak:2006ab}. 

\newsection{Conclusions \label{sec:concl}}

We have presented details of our large-$m_c$
calculation~\cite{Misiak:2006ab} of those NNLO corrections to\linebreak
${\cal B}(\bar B \to X_s \gamma)$~ that still require interpolation in
$m_c$. Applying Lagrangian-level decoupling rather than the purely
diagrammatic asymptotic expansions has led to appreciable
simplifications of the analysis.

Our results are going to be useful again in the near future when the
calculation of $K_{17}^{(2)\rm rem}$ and $K_{27}^{(2)\rm rem}$ at
$m_c=0$ is completed~\cite{Czakon:2010xxx} providing data for an
upgraded interpolation in $m_c$. With those inputs, as well as new
results for $K_{78}^{(2)\rm rem}$~\cite{Asatrian:2010xxx} and the
remaining BLM terms~\cite{Misiak:2010xxx}, an update of the
phenomenological analysis~\cite{Misiak:2006zs,Misiak:2006ab} will be
mandatory.  An ultimate goal is to 
\linebreak\newpage\noindent
make the perturbative uncertainties in~ ${\cal B}(\bar B\to X_s\gamma)$~
negligible with respect to the non-perturbative~\cite{Benzke:2010js} and
experimental~\cite{Barberio:2008fa} ones.

\section*{Acknowledgments}

This work has been supported by the DFG through SFB/TR9 ``Computational
Particle Physics'' and the ``Mercator'' guest professorship programme.
M.M. acknowledges partial support from the EU-RTN Programme
``FLAVIAnet'' (MRTN-CT-2006-035482), and from the Polish Ministry of
Science and Higher Education as a research project N~N202~006334 (in
years 2008-11).  We thank the Galileo Galilei Institute for Theoretical
Physics for hospitality and the INFN for partial support during
completion of this work.

\newappendix{Appendix}
\def\theequation{A.\arabic{equation}}

Here, we quote explicit expressions for the relevant functions 
$\phi_{ij}^{(1)}(\delta)$~\cite{Pott:1995if,Ali:1990tj}:
\bea
\phi^{(1)}_{77} &=& -\fm{2}{3} \ln^2 \delta -\fm{7}{3} \ln \delta - \fm{31}{9} + \fm{10}{3} \delta + \fm{1}{3} \delta^2 
                    -\fm{2}{9} \delta^3 + \fm{1}{3} \delta ( \delta - 4 ) \ln \delta,\\[1mm]
\phi^{(1)}_{78} &=& \fm{8}{9} \left[ {\rm Li}_2(1-\delta) - \fm{1}{6}\pi^2 - \delta \ln \delta + \fm{9}{4} \delta 
                    - \fm{1}{4} \delta^2 + \fm{1}{12} \delta^3 \right],\\[1mm]
\phi^{(1)}_{88} &=& \fm{1}{27} \left\{ \left[ \delta^2 + 2 \delta + 4 \ln(1-\delta) \right] 
                    \ln \f{m_s^2}{m_b^2} \right. \nnb\\
                &+& \left. 4{\rm Li}_2(1-\delta) -\fm{2}{3} \pi^2 -\delta(2+\delta)\ln\delta + 8\ln(1-\delta) 
                    - \fm{2}{3} \delta^3 + 3 \delta^2 + 7 \delta \right\}.\label{phi88} \\[1mm]
\phi_{47}^{(1)}(\delta) &=& -3 \phi_{48}^{(1)}(\delta) ~=~
-\fm{1}{54} \delta \left( 1 - \delta + \fm{1}{3} \delta^2 \right)~
-~ \fm{1}{4}~ \lim_{m_c \to m_b} \phi_{27}^{(1)}(\delta)\nnb\\[1mm]
&=& \fm{1}{54} \pi \left( 3 \sqrt{3} - \pi \right) 
  + \fm{1}{81}\delta^3 - \fm{25}{108} \delta^2 + \fm{5}{54} \delta
  + \fm{2}{9} \left( \delta^2 + 2\delta + 3 \right) \arctan^2\sqrt{\f{1-\delta}{3+\delta}}\nnb\\[1mm]
&-& \fm{1}{3} \left( \delta^2 + 4\delta + 3 \right) \sqrt{\f{1-\delta}{3+\delta}}\, \arctan\sqrt{\f{1-\delta}{3+\delta}}.
\label{phi47}
\eea
The functions $\phi_{47}^{(1)}$ and $\phi_{48}^{(1)}$ are exactly the
same in the 5-flavour and 4-flavour theories.\footnote{
$\;$Eq.~(3.12) of Ref.~\cite{Misiak:2006ab} contains a misprint in the
coefficient at~ $\lim_{m_c \to m_b}$~ which we correct in the first line
of Eq.~(\ref{phi47}) here.}
They are generated by the $s$- and $b$-quark loops with no Dirac traces
only.  Contributions with traces cancel out in the same way as in the QED
electron-loop contributions to three-photon interactions (Furry theorem).

Let us note that $\phi^{(1)}_{88}$ in Eq.~(\ref{phi88}),
$\phi_{88}^{(2)\beta_0}$, as well as the neglected diagrams in
Fig.~\ref{fig:bsqqgam} contain collinear logarithms where~ $m_q\neq 0$~
for~ $q=u,d,s$~ need to be retained at the perturbative level. The
actual collinear regulators in reality are of order of the light meson
masses ($m_\pi$, $m_K$).  Non-perturbative collinear effects in $\bar B
\to X_s \gamma$ have been discussed in
Refs.~\cite{Benzke:2010js,Kapustin:1995fk}.  Their numerical effect on
the branching ratio for $E_0 \sim \fm{m_b}{3}$ is generically small
($\sim 1\%$) thanks to the phase-space suppression, electric charge
factors and/or small values of the relevant WCs.

\setlength {\baselineskip}{0.2in}
 

\begin{thebibliography}{99}
%
\bibitem{Olive:2008vv}
  See, e.g., K.~A.~Olive and L.~Velasco-Sevilla,
  %``Constraints on Supersymmetric Flavour Models from b->s gamma,''
  JHEP {\bf 0805} (2008) 052
  [arXiv:0801.0428].
  %%CITATION = JHEPA,0805,052;%%
%
\bibitem{Benzke:2010js}
  M.~Benzke, S.~J.~Lee, M.~Neubert and G.~Paz,
  %``Factorization at Subleading Power and Irreducible Uncertainties in $\bar B\to X_s\gamma$ Decay,''
  arXiv:1003.5012.
  %%CITATION = ARXIV:1003.5012;%%
%
\bibitem{Misiak:2009nr}
  M.~Misiak,
  %``B -> Xs gamma - Current Status,''
  Acta Phys.\ Polon.\  B {\bf 40} (2009) 2987
  [arXiv:0911.1651].
  %%CITATION = APPOA,B40,2987;%%
%
\bibitem{Misiak:2006zs}
  M.$\,$Misiak {\it et al.},
  %``The first estimate of B(anti-B --> X/s gamma) at O(alpha(s)**2),''
  Phys.\ Rev.\ Lett.\  {\bf 98} (2007) 022002
  [hep-ph/0609232].
  %%CITATION = PRLTA,98,022002;%%
%
\bibitem{Misiak:2006ab}
  M.~Misiak and M.~Steinhauser,
  % ``NNLO QCD corrections to the B -> X_s gamma matrix elements using interpolation in m_c,''
  Nucl.\ Phys.\ B {\bf 764} (2007) 62 
  [hep-ph/0609241].
  %%CITATION = HEP-PH 0609241;%%
%
\bibitem{Barberio:2008fa}
  E.$\,$Barberio {\it et al.} (Heavy Flavour Averaging Group),
  %``Averages of b-hadron and c-hadron Properties at the End of 2007,''
  arXiv:0808.1297, and on-line update at
  {\tt http://www.slac.stanford.edu/xorg/hfag/rare/winter10/radll/btosg.pdf}  
  %%CITATION = ARXIV:0808.1297;%%
%
\bibitem{Artuso:2009jw}
  M.$\,$Artuso, E.$\,$Barberio and S.$\,$Stone,
  %``$B$ Meson Decays,''
  PMC~Phys.$\,$A~{\bf 3}~(2009)~3 
  [arXiv:0902.3743].
  %%CITATION = PMCPA,A3,3;%%
%
\bibitem{Chen:2001fj} 
  S.$\,$Chen {\it et al.} (CLEO Collaboration),
  %``Branching fraction and photon energy spectrum for b $\to$ s gamma,''
  Phys.\ Rev.\ Lett.\  {\bf 87} (2001) 251807
  [hep-ex/0108032].
  %%CITATION = HEP-EX 0108032;%%
%
\bibitem{Aubert:2005cu}
  B.$\,$Aubert {\it et al.} (BABAR Collaboration),
  %``Measurements of the $B \to X_s \gamma$ branching fraction and photon
  %spectrum from a sum of exclusive final states,''
  Phys.\ Rev.\ D {\bf 72} (2005) 052004  
  [hep-ex/0508004],
  %%CITATION = HEP-EX 0508004;%%
%
% \bibitem{Aubert:2006gg}
% B.$\,$Aubert {\it et al.}  (BABAR Collaboration),
  %``Measurement of the branching fraction and photon energy moments of B  -->
  %X/s gamma and A(CP)(B --> X(s+d) gamma),''
  Phys.\ Rev.\ Lett.\  {\bf 97} (2006) 171803
  [hep-ex/0607071],
  %%CITATION = PRLTA,97,171803;%%
%
% \bibitem{Aubert:2007my}
% B.$\,$Aubert {\it et al.}  (BABAR Collaboration),
  %``Measurement of the B --> X_s gamma Branching Fraction and Photon Energy
  %Spectrum using the Recoil Method,''
  Phys.\ Rev.\  D {\bf 77} (2008) 051103
  [arXiv:0711.4889].
  %%CITATION = PHRVA,D77,051103;%%
%
\bibitem{Abe:2001hk}
  K.$\,$Abe {\it et al.} (BELLE Collaboration),
  %``A measurement of the branching fraction for the inclusive B $\to$ X/s  gamma decays with BELLE,''
  Phys.\ Lett.\ B {\bf 511} (2001) 151
  [hep-ex/0103042].
  %%CITATION = HEP-EX 0103042;%%
%
\bibitem{Limosani:2009qg}
  A.~Limosani {\it et al.}  (Belle Collaboration),
  %``Measurement of Inclusive Radiative B-meson Decays with a Photon Energy
  %Threshold of 1.7 GeV,''
  Phys.\ Rev.\ Lett.\  {\bf 103} (2009) 241801
  [arXiv:0907.1384].
  %%CITATION = PRLTA,103,241801;%%
%
\bibitem{Brodsky:1982gc}
  S.~J.~Brodsky, G.~P.~Lepage and P.~B.~Mackenzie,
  %``On The Elimination Of Scale Ambiguities In Perturbative Quantum
  %Chromodynamics,''
  Phys.\ Rev.\ D {\bf 28} (1983) 228. 
  %%CITATION = PHRVA,D28,228;%%
%
\bibitem{Misiak:2004ew}
  M.$\,$Misiak and M.$\,$Steinhauser,
  %``Three-loop matching of the dipole operators for b $\to$ s gamma and b $\to$ s g,''
  Nucl.\ Phys.\ B {\bf 683} (2004) 277 
  [hep-ph/0401041].
  %%CITATION = HEP-PH 0401041;%%
%
 \bibitem{Gorbahn:2004my}
  M.$\,$Gorbahn and U.$\,$Haisch,
  %``Effective Hamiltonian for non-leptonic $|$Delta(F)$|$ = 1 decays at NNLO in QCD,''
  Nucl.\ Phys.\ B {\bf 713} (2005) 291
  [hep-ph/0411071].
  %%CITATION = HEP-PH 0411071;%%
%
\bibitem{Bobeth:1999mk}
  C.$\,$Bobeth, M.$\,$Misiak and J.$\,$Urban,
  %``Photonic penguins at two loops and m(t)-dependence of BR(B $\to$ X(s) l+  l-),''
  Nucl.\ Phys.\ B {\bf 574} (2000) 291
  [hep-ph/9910220];
  %%CITATION = HEP-PH 9910220;%%
%\bibitem{Gorbahn:2005sa}
  M.$\,$Gorbahn, U.$\,$Haisch and M.$\,$Misiak,
  %``Three-loop mixing of dipole operators,''
  Phys.\ Rev.\ Lett.\  {\bf 95} (2005) 102004
  [hep-ph/0504194];
  %%CITATION = HEP-PH 0504194;%%
%\bibitem{Czakon:2006ss}
  M.$\,$Czakon, U.$\,$Haisch and M.$\,$Misiak,
  %''Four-loop anomalous dimensions for radiative flavour-changing decays''
  JHEP {\bf 0703} (2007) 008
  [hep-ph/0612329].
  %%CITATION = HEP-PH 0612329;%%
%
\bibitem{Bieri:2003ue}
K.~Bieri, C.~Greub and M.~Steinhauser,
%``Fermionic NNLL corrections to b $\to$ s gamma,''
Phys.\ Rev.\ D {\bf 67} (2003) 114019 
[hep-ph/0302051].
%%CITATION = HEP-PH 0302051;%%
%
\bibitem{Boughezal:2007ny}
    R.~Boughezal, M.~Czakon and T.~Schutzmeier,
    %``NNLO fermionic corrections to the charm quark mass dependent matrix
    %elements in B -> X_s gamma,''
    JHEP {\bf 0709}, 072 (2007)
    [arXiv:0707.3090].
    %%CITATION = JHEPA,0709,072;%%
%
\bibitem{Ligeti:1999ea}
  Z.~Ligeti, M.E.~Luke, A.V.~Manohar and M.B.~Wise,
  %``The anti-B $\to$ X/s gamma photon spectrum,''
  Phys.\ Rev.\ D {\bf 60}, 034019 (1999)
  [hep-ph/9903305].
  %%CITATION = HEP-PH 9903305;%%
%
\bibitem{Ferroglia:2010xxx}
  A.~Ferroglia, P.~Gambino and U.~Haisch, to be published.
%
\bibitem{Misiak:2010xxx}
  M.~Misiak and M.~Poradzi\'nski, in preparation.
%
\bibitem{Asatrian:2006rq}
  H.~M.~Asatrian, T.~Ewerth, H.~Gabrielyan and C.~Greub,
  %``Charm quark mass dependence of the electromagnetic dipole operator
  %contribution to anti-B --> X/s gamma at O(alpha(s)**2),''
  Phys.\ Lett.\  B {\bf 647}, 173 (2007)
  [hep-ph/0611123];
  %%CITATION = PHLTA,B647,173;%%
%\bibitem{Ewerth:2008nv}
  T.~Ewerth,
  %``Fermionic corrections to the interference of the electro- and
  %chromomagnetic dipole operators in anti-B --> X(s) gamma at O(alpha(s)**2),''
  Phys.\ Lett.\  B {\bf 669}, 167 (2008)
  [arXiv:0805.3911].
  %%CITATION = PHLTA,B669,167;%%
%
\bibitem{Smi02}
V.~A. Smirnov,
{\it Applied Asymptotic Expansions in Momenta and Masses},
Springer-Verlag, Heidelberg, 2001.
%
\bibitem{Buras:2002tp}
  A.J.~Buras, A.~Czarnecki, M.~Misiak and J.~Urban,
  %``Completing the NLO QCD calculation of anti-B $\to$ X/s gamma,''
  Nucl.\ Phys.\ B {\bf 631}, 219 (2002)
  [hep-ph/0203135].
  %%CITATION = HEP-PH 0203135;%%
%
\bibitem{Harlander:1997zb}
  R.~Harlander, T.~Seidensticker and M.~Steinhauser,
  %``Complete corrections of O(alpha alpha(s)) to the decay of the Z
  %boson  into bottom quarks,''
  Phys.\ Lett.\ B {\bf 426} (1998) 125,
  hep-ph/9712228.
  %%CITATION = HEP-PH 9712228;%%
%
\bibitem{Seidensticker:1999bb}
  T.~Seidensticker,
  %``Automatic application of successive asymptotic expansions of Feynman diagrams,''
  hep-ph/9905298.
  %%CITATION = HEP-PH 9905298;%%
%
\bibitem{Steinhauser:2000ry}
  M.~Steinhauser,
  %``MATAD: A program package for the computation of massive tadpoles,''
  Comput.\ Phys.\ Commun.\  {\bf 134} (2001) 335,
  hep-ph/0009029.
  %%CITATION = CPHCB,134,335;%%
%
\bibitem{Fleischer:1999tu}
  J.~Fleischer and M.~Y.~Kalmykov,
  %``ON-SHELL2: FORM based package for the calculation of two-loop
  %self-energy single scale Feynman diagrams occurring in the standard  model,''
  Comput.\ Phys.\ Commun.\  {\bf 128} (2000) 531
  [hep-ph/9907431].
  %%CITATION = CPHCB,128,531;%%
%
\bibitem{Anastasiou:2004vj}
  C.~Anastasiou and A.~Lazopoulos,
  %``Automatic integral reduction for higher order perturbative calculations,''
  JHEP {\bf 0407} (2004) 046
  [hep-ph/0404258].
  %%CITATION = JHEPA,0407,046;%%
%
\bibitem{Laporta:2001dd}
  S.~Laporta,
  %``High-precision calculation of multi-loop Feynman integrals by  difference equations,''
  Int.\ J.\ Mod.\ Phys.\  A {\bf 15} (2000) 5087
  [hep-ph/0102033].
  %%CITATION = IMPAE,A15,5087;%%
%
\bibitem{Smirnov:2008iw}
  A.~V.~Smirnov,
  %``Algorithm FIRE -- Feynman Integral REduction,''
  JHEP {\bf 0810} (2008) 107
  [arXiv:0807.3243].
  %%CITATION = JHEPA,0810,107;%%
%
\bibitem{Greub:1996jd}
  C.~Greub, T.~Hurth and D.~Wyler,
  %``Virtual corrections to the decay $b \to s \gamma$,''
  Phys.\ Lett.\ B {\bf 380}, 385 (1996)
  [hep-ph/9602281],
  %%CITATION = HEP-PH 9602281;%%
  %\bibitem{Greub:1996tg}
  % C.~Greub, T.~Hurth and D.~Wyler,
  %``Virtual $O(\a_s)$ corrections to the inclusive decay $b \to s \gamma$,''
  Phys.\ Rev.\ D {\bf 54}, 3350 (1996)
  [hep-ph/9603404];
  %%CITATION = HEP-PH 9603404;%%
  %\bibitem{Buras:2001mq}
  A.~J.~Buras, A.~Czarnecki, M.~Misiak and J.~Urban,
  %``Two-loop matrix element of the current-current operator in the decay b
  %$\to$ X/s gamma,''
  Nucl.\ Phys.\ B {\bf 611}, 488 (2001)
  [hep-ph/0105160].
  %%CITATION = HEP-PH 0105160;%%
%
\bibitem{Politzer:1980me}
  H.~D.~Politzer,
  %``Power Corrections At Short Distances,''
  Nucl.\ Phys.\  B {\bf 172} (1980) 349.
  %%CITATION = NUPHA,B172,349;%%
%
\bibitem{Chetyrkin:1997un}
  K.~G.~Chetyrkin, B.~A.~Kniehl and M.~Steinhauser,
  %``Decoupling relations to O(alpha(s)**3) and their connection to  low-energy
  %theorems,''
  Nucl.\ Phys.\  B {\bf 510} (1998) 61
  [hep-ph/9708255].
  %%CITATION = NUPHA,B510,61;%%
%
\bibitem{Grozin:2007fh}
  A.~G.~Grozin, P.~Marquard, J.~H.~Piclum and M.~Steinhauser,
  %``Three-Loop Chromomagnetic Interaction in HQET,''
  Nucl.\ Phys.\  B {\bf 789} (2008) 277
  [arXiv:0707.1388].
  %%CITATION = NUPHA,B789,277;%%
%
\bibitem{Pott:1995if}
  N.~Pott,
  %``Bremsstrahlung corrections to the decay $b \to s \gamma$,''
  Phys.\ Rev.\ D {\bf 54}, 938 (1996)
  [hep-ph/9512252].
  %%CITATION = HEP-PH 9512252;%%
%
\bibitem{Czakon:2010xxx}
  R.~Boughezal, M.~Czakon and T.~Schutzmeier, in preparation;
  M.~Czakon, T.~Huber and T.~Schutzmeier, in preparation.
%
\bibitem{Asatrian:2010xxx}
  H.~Asatrian, T.~Ewerth, A.~Ferroglia, C.~Greub and G.~Ossola, in preparation.
%
\bibitem{Ali:1990tj}
  A.~Ali and C.~Greub,
  %``Inclusive Photon Energy Spectrum In Rare B Decays,''
  Z.\ Phys.\ C {\bf 49}, 431 (1991);
  %%CITATION = ZEPYA,C49,431;%%
  %\bibitem{Ali:1990vp}  
  % A.~Ali and C.~Greub,  
  %``A Profile Of The Final States In B $\to$ X(S) Gamma And An Estimate Of The Branching Ratio Br (B $\to$ K* Gamma),''
  Phys.\ Lett.\ B {\bf 259}, 182 (1991);
  %%CITATION = PHLTA,B259,182;%%  
  %\bibitem{Ali:1995bi}  
  % A.~Ali and C.~Greub,  
  %``Photon energy spectrum in B $\to$ X(s) + gamma and comparison with data,''  
  Phys.\ Lett.\ B {\bf 361}, 146 (1995)
  [hep-ph/9506374].
  %%CITATION = HEP-PH 9506374;%%
%
\bibitem{Kapustin:1995fk}
  A.~Kapustin, Z.~Ligeti and H.~D.~Politzer,
  %``Leading Logarithms Of The B Quark Mass In Inclusive B $\to$ X(S) Gamma
  %Decay,''
  Phys.\ Lett.\  B {\bf 357} (1995) 653
  [hep-ph/9507248].
  %%CITATION = PHLTA,B357,653;%%
%
\end{thebibliography}
\end{document}